\documentclass[prb,aps,twocolumn]{revtex4}
\usepackage{amssymb}
\usepackage{graphicx}
\usepackage{amsmath}

\begin{document}

\title{Magnetic-field-induced dimensional crossover in the organic metal $%
\alpha $-(BEDT-TTF)$_{2}$KHg(SCN)$_{4}$ }
\author{P. D. Grigoriev$^{1}$}
\author{M. V. Kartsovnik$^{2}$}
\author{W. Biberacher$^{2}$}
\affiliation{$^{1}$ L. D. Landau Institute for Theoretical Physics, Russian Academy of
Sciences, 142432 Chernogolovka, Russia}
\affiliation{$^{2}$Walther-Meissner-Institut, Bayerische Akademie der Wissenschaften,
D-85748 Garching, Germany}
\date{\today }

\begin{abstract}
The field dependence of interlayer magnetoresistance of the pressurized (to
the normal state) layered organic metal $\alpha $-(BEDT-TTF)$_{2}$KHg(SCN)$%
_{4}$ is investigated. The high quasi-two-dimensional anisotropy, when the
interlayer hopping time is longer than the electron mean-free time and than
the cyclotron period, leads to a dimensional crossover and to strong
violations of the conventional three-dimensional theory of
magnetoresistance. The monotonic field dependence is found to change from
the conventional behavior at low magnetic fields to an anomalous one at high
fields. The shape of Landau levels, determined from the damping of magnetic
quantum oscillations, changes from Lorentzian to Gaussian. This indicates
the change of electron dynamics in the disorder potential from the usual
coherent three-dimensional regime to a new regime, which can be referred to
as weakly coherent.
\end{abstract}

\maketitle

\section{Introduction}

Highly anisotropic layered conductors in a strong magnetic field
may undergo a dimensional crossover from three-dimensional (3D) to
almost two-dimensional (2D) electron dynamics when the interlayer
transfer integral $t_{\perp }$ becomes smaller than the other
relevant parameters: the scattering rate and the cyclotron
frequency. This crossover may change the electronic transport
properties in various layered compounds: organic metals,
heterostructures, intercalated compounds, superconductive cuprates
and pnictides, cobaltates etc. Scattering by crystal disorder is
one of the most frequently discussed mechanisms of breaking the
interlayer band transport in layered metals. If the scattering
rate $\tau ^{-1}$ is larger than the interlayer hopping rate,
$\tau _{h}^{-1}\sim t_{\perp }/\hbar $, the quasiparticle momentum
and Fermi surface are only defined within conducting layers, i.e.
become strictly 2D. However, the interlayer electron tunneling may
still be "coherent" and conserve the in-plane electron momentum.
The corresponding regime was called "weakly
incoherent".\cite{kenz98,MosesMcKenzie1999} In the literature
there has been a long-time discussion, supported by
theoretical\cite{MosesMcKenzie1999,Abrikosov1999,Lundin2003,OsadaIncoherent,Ho,Gvozd2007,McKenzie2007,Maslov,WIPRB2011,WIJETP2011,WIFNT2011,Incoh2009}
 and experimental\cite{Incoh2009,Zuo1999,Wosnitza2002,CrossoverNAture2002,MarkPRL2006,Ardavan2006,zver96}
arguments, whether this "weakly incoherent" regime can be distinguished from
the usual three-dimensional (3D) electron transport. Up to now, no
considerable qualitative differences between 3D and "weakly incoherent"
regimes have been suggested or observed. The only significant predicted
change in the "weakly incoherent" regime is the absence of the narrow
"coherence peak" on the angular dependence of magnetoresistance when the
magnetic field is directed along the conducting layers\cite%
{kenz98,MosesMcKenzie1999}. However, the absence of even this subtle feature
in the "weakly incoherent" regime has not received a sound proof yet. Hence,
for a long time it was believed\cite%
{kenz98,MosesMcKenzie1999,McKenzie2007,Maslov,kuma92} that in this regime
the interlayer resistivity $\rho _{\perp }(T)$ is nearly identical to that
in the fully coherent three-dimensional (3D) case.

Another possible mechanism of dimensional crossover is associated with
external magnetic field $B$.\cite{CommentT} Indeed, the behavior of magnetic
quantum oscillations (MQO) is substantially modified\cite%
{Shoenberg,PhSh,Shub,SO,ChampelMineev,Gvozd2007,harr96,mk04} when the
cyclotron frequency $\omega _{c}=eB/m_c$ becomes larger than the interlayer
hopping rate $\tau_{h}^{-1}$. However, the existing theories predict no
significant changes in the electron dynamics at weak (but coherent)
interlayer electron hopping unless additional mechanisms of interlayer
electron transport such as interlayer hopping via the resonance
impurities\cite{Abrikosov1999,Maslov,Incoh2009} or
boson-assisted\cite{Lundin2003,Ho} tunneling are
concerned.\cite{commentFISDW}

In this work we show, that the parameter $b_{\ast }\equiv \hbar \omega
_{c}/t_{\perp }$ drives a transition between two qualitatively different
regimes of electron dynamics. There are several principal distinctions in
the field dependence of interlayer magnetoresistance at $b_{\ast }\gg 1$,
originating from the qualitatively different influence of disorder on
electronic properties. These changes cannot be explained by a simple
extension of the formulas in Refs. \cite{Shub,ChampelMineev} to the case
$b_{\ast }\gg 1$, which unambiguously separates this regime from
$b_{\ast }\ll 1$. We refer to this new regime as "\textit{weakly coherent}":
it implies a conservation of the in-plane electron momentum by the
interlayer tunneling term in the Hamiltonian; on the other hand, the time
scale of this tunneling is much larger than the cyclotron period. Note
that the parameter $b_{\ast }\equiv \hbar \omega _{c}/t_{\perp }$ is
completely different from the parameter $\hbar /t_{\perp }\tau $, which was
used\cite{kenz98,MosesMcKenzie1999} to separate the coherent and weakly
incoherent regimes. The proposed weakly coherent regime imposes no
limitation on the value of parameter $\hbar /t_{\perp }\tau$. Therefore,
strictly speaking, there is no direct relation between the weakly incoherent
and the newly defined weakly coherent regimes. On the other hand, the
compounds satisfying the condition of the weakly incoherent regime are
automatically driven to the weakly coherent regime by a strong magnetic
field such that $\omega_c\tau >1$.

Here we present a joint theoretical and experimental study of the weakly
coherent regime, on the example of the layered organic metal
$\alpha $-(BEDT-TTF)$_{2}$KHg(SCN)$_{4}$. The title compound has a
strong anisotropy with the interlayer transfer integral
$t_{\perp }\sim 30$\;$\mu$eV.\cite{MarkPRL2006}
At ambient pressure it undergoes a charge-density-wave
transition at $\approx 8.5$\;K, which can be suppressed by applying an
external pressure $P>P_c\approx 2.5$\;kbar.\cite{andres01,andres05}. To
avoid complications related to the zero-field and magnetic-field-induced
charge-density-wave states \cite{andres12}, a pressure of 6\;kbar,
considerably exceeding $P_c$ and temperatures above 1\;K were used, so that
the compound was in the fully normal metallic state in our experiment. The
corresponding Fermi surface consists of a cylinder (quasi-2D band) and a
pair of weakly warped open sheets (quasi-1D band). As will be shown below,
the interlayer conductivity in fields above 2~T, is largely determined by
the quasi-2D carriers, so we will restrict our analysis to the case of a
cylindrical Fermi surface. The weakly-coherent criterion
$b_{\ast}\gg 1$ in this compound is satisfied at an easily accessible field
$B_{z}\gg t_{\perp}m_c/e\hbar \simeq 0.3$\;T, where
$m_c \approx 1.3m_{\mathrm{e}}$ is the relevant cyclotron mass.

\section{Theoretical background}

The first step in the theoretical analysis of the weakly coherent regime was
made recently in Refs. \cite{WIPRB2011,WIJETP2011,WIFNT2011}, where it was
shown that in the regime where weakly coherent and weakly incoherent
criteria overlap, the earlier theoretical conclusion\cite%
{kenz98,MosesMcKenzie1999,McKenzie2007} that the interlayer resistivity
$\rho _{\perp }(T)$ is identical to that in the fully coherent
three-dimensional (3D) case is not valid. The new analysis, going beyond the
constant relaxation time approximation used in the earlier
works,\cite{kenz98,MosesMcKenzie1999,ChampelMineev,harr96} has predicted
several qualitatively new features of interlayer magnetoresistance in the
weakly coherent regime.

The first prediction is a monotonic growth of the magnetoresistance, averaged
over MQO, with an increase of magnetic field, parallel to the current and
perpendicular to the conducting layers.\cite{WIPRB2011,WIJETP2011,WIFNT2011}
This increase, contradicting the classical theory of MR even for quasi-2D
metals,\cite{Shub,ChampelMineev} is due to the enhancement of the effect of
short-range impurities caused by a magnetic field and follows directly from
the monotonic growth $\propto \sqrt{B_{z}}$ of the Landau level (LL)
broadening due to the short-range impurity scattering.\cite{Ando} It is not
related to the low crystal symmetry. The field dependence of the
nonoscillating component of the interlayer conductivity is given
by \cite{WIPRB2011,WIJETP2011,WIFNT2011}
\begin{equation}
\bar{\sigma}_{zz}\left( B\right) \approx \sigma _{0}\left[ \left(
\alpha\omega _{c}\tau \right) ^{2}+1\right] ^{-1/4}.  \label{sSS}
\end{equation}
The numerical coefficient $\alpha \approx 2$ before $\omega _{c}\tau $ is
not universal and slightly depends on the shape of LLs.\cite{WIJETP2011}

The second prediction for the weakly coherent regime\cite{WIPRB2011} is a
modification of the angular dependence of magnetoresistance due to a
decrease of the effective mean scattering time $\tau $ with an increase of
the interlayer component $B_{z}=B\cos \theta $ of magnetic field [see Eqs.
(36) and (37) of Ref. \cite{WIPRB2011}]. An accurate comparison of this
effect with experiment on $\alpha $-(BEDT-TTF)$_{2}$KHg(SCN)$_{4}$ requires
elimination of the angular dependence associated with the quasi-1D parts of
the Fermi surface which is beyond the scope of this work.

The third prediction of the theory in
Refs. \cite{WIPRB2011,WIJETP2011,WIFNT2011}, is the growth of the Dingle
temperature of MQO with an increase of magnetic field, and, hence, the
stronger damping of
MQO. Naively, since the LL width $\Gamma \equiv \hbar /2\tau _{B}$ in the
single-site approximation\cite{Ando} grows at $\omega _{c}\tau \gg 1$ as $%
\tau /\tau _{B}\approx \left[ \left( 2\omega _{c}\tau \right) ^{2}+1\right]
^{1/4}$ $\propto \sqrt{B_{z}}$, one would expect the similar square-root
growth of the Dingle temperature $T_{D}\left( B_{z}\right) $. However, this
simple conclusion is incorrect for two reasons: (i) the square-root growth
of the LL width appears only for a short-range impurity potential, while in
organic and many other layered metals the main contribution to the LL
broadening often comes from a long-range disorder potential\cite{SO};
(ii) the MQO damping factor is determined not only by the width of LLs,
but also by their shape.

To check this we substitute the density of state (DoS)
\begin{equation}
\rho \left( \varepsilon \right) =\sum_{n\geq 0}D\left[ \varepsilon -\hbar
\omega _{c}\left( n+1/2\right) \right]  \label{rhoDoS}
\end{equation}%
to the expression for the interlayer conductivity, obtained as a linear
response from the Kubo formula [see Eq. (14) of Ref. \cite{WIJETP2011} and
note that $\rho \left( \varepsilon \right) =-$Im$G_{R}(\varepsilon )/\pi $]
\begin{equation}
\sigma _{zz}=\pi \sigma _{0}\Gamma _{0}\hbar \omega _{c}\sum_{s=\uparrow
,\downarrow }\int d\varepsilon \left[ -n_{F}^{\prime }(\varepsilon )\right]
\left\vert \rho _{s}\left( \varepsilon \right) \right\vert ^{2}.  \label{sp1}
\end{equation}%
As long as the shape and width of LLs do not change with temperature, the
temperature harmonic damping factor $R_{T}$ is described by the usual
Lifshitz-Kosevich expression: $R_{T}\left( k\right) =kX/\sinh \left(
kX\right) \,$, where $X\equiv 2\pi ^{2}k_{B}T/\hbar \omega _{c}$. Now
substituting Eq. (\ref{rhoDoS}) to Eq. (\ref{sp1}) and applying the Poisson
summation formula, we obtain at $N_{\mathrm{LL}}\gg 1$%
\begin{equation}
\frac{\sigma _{zz}}{\bar{\sigma}_{zz}}=\sum_{k=-\infty }^{\infty }\left(
-1\right) ^{k}\exp \left( \frac{2\pi ik\mu }{\hbar \omega _{c}}\right) R_{%
\mathrm{D}}\left( k\right) R_{T}\left( k\right) R_{\mathrm{S}}\left(
k\right) ,  \label{s1}
\end{equation}%
where the averaged over MQO interlayer conductivity $\bar{\sigma}_{zz}$ is
given by Eq. (\ref{sSS}), the spin-splitting damping factor\cite{Shoenberg} $%
R_{\mathrm{S}}\left( k\right) =\cos \left( \pi kgm^{\ast }/2\right) $, $\mu $
is the Fermi energy, $m^{\ast }\equiv m_{c}/m_{\mathrm{e}}$ is the effective
cyclotron mass normalized to the free electron mass, and the Dingle factor%
\begin{equation}
R_{\mathrm{D}}\left( k\right) =2\pi \Gamma \int_{-\infty }^{\infty }\exp
\left( \frac{2\pi ikE}{\hbar \omega _{c}}\right) \left\vert D\left( E\right)
\right\vert ^{2}dE.  \label{RD}
\end{equation}

The traditional Lorentzian shape of LLs with the halfwidth $\Gamma $, $D_{%
\mathrm{L}}\left( E\right) =\left( \pi \Gamma \right) ^{-1}/\left[ 1+\left(
E/\Gamma \right) ^{2}\right] $, after substitution to Eq. (\ref{s1}) gives
the Dingle factor%
\begin{equation}
R_{\mathrm{DL}}\left( k\right) =\exp \left( -2\pi k\Gamma /\hbar \omega
_{c}\right) \left( 1+2\pi k\Gamma /\hbar \omega _{c}\right) .  \label{RDLM}
\end{equation}%
As was shown in Refs. \cite{ShubCondMat,ChampelMineev,Shub}, it
differs from the standard Dingle factor, valid in the case
$t_{\perp} \gg \hbar\omega_c$,
\begin{equation}
R_{\mathrm{DL}}\left( k\right) \approx \exp \left( -2\pi k\Gamma /\hbar
\omega _{c}\right) .  \label{RDL}
\end{equation}
However, this difference does not considerably change the Dingle plot, i.e.
the field dependence of the logarithm of the Dingle factor:
\begin{eqnarray}
\ln R_{\mathrm{DL}} &=&-2\pi \Gamma /\hbar \omega _{c}+\ln \left( 1+2\pi
k\Gamma /\hbar \omega _{c}\right)  \notag \\
&=&-B_{0}/B+\ln \left( 1+B_{0}/B\right) ,  \label{DinglePlot}
\end{eqnarray}
where $B_{0}=2\pi k\Gamma m_c/\hbar e$.
In a strong field, when $\omega _{c}\tau \gg 1$ the ratio $B_0/B$ is small
and the correction $\ln \left( 1+B_{0}/B\right) \ll 1$. In the
opposite limit, $\omega _{c}\tau \ll 1$ or $B\ll B_{0}$, the field dependence
coming from the first term in Eq. (\ref{DinglePlot}) is much stronger than
weak logarithmic dependence from the second term. Hence,
the factor $(1+2\pi k\Gamma /\hbar \omega _{c})$ gives only a small
correction to the field dependence of the MQO amplitude
[see Fig. \ref{FigDingle1} below for comparison of Eqs. (\ref{RDLM})
and (\ref{RDL}), and one usually can apply Eq. (\ref{RDL}) for the
analysis of the Dingle plots.

If one assumes $\Gamma $ to be independent of $B$, Eq. (\ref{RDL}) gives the
standard result:

\begin{equation}
R_{\mathrm{DL}}\left( k\right) \approx \exp \left( -\mathrm{const}\cdot
k/B_{z}\right) ,  \label{RDL1}
\end{equation}%
while if $\Gamma \propto \sqrt{B_{z}}$ as in the self-consistent Born
approximation,\cite{Ando} Eq. (\ref{RDL}) gives%
\begin{equation}
R_{\mathrm{DL}}^{\ast }\left( k\right) \approx \exp \left( -\mathrm{const}%
\cdot k/\sqrt{B_{z}}\right) .  \label{RDL2}
\end{equation}

The Gaussian shape of LLs, $D_{\mathrm{G}}\left( E\right) =\left( \sqrt{\pi }%
\Gamma \right) ^{-1}\exp \left( -E^{2}/\Gamma ^{2}\right) $, gives the
Dingle factor
\begin{equation}
R_{\mathrm{DG}}\left( k\right) =\sqrt{\pi /2}\exp \left[ -\left( \pi k\Gamma
/\hbar \omega _{c}\right) ^{2}/2\right] .  \label{RDG}
\end{equation}%
The theory predicts the Gaussian shape of the Landau levels (for a review
see, e.g., Ref. \cite{KukushikUFN1988}) for a physically reasonable
white-noise or Gaussian correlator of the disorder potential $U\left(
\mathbf{r}\right) $:%
\begin{equation}
Q\left( \mathbf{r}\right) =\left\langle U\left( \mathbf{0}\right) U\left(
\mathbf{r}\right) \right\rangle \propto \exp \left( -r^{2}/2d^{2}\right) .
\label{Q}
\end{equation}%
For a long-range disorder potential, when $d\gg l_{B}\equiv \sqrt{\hbar /eB}$%
, the LL width $\Gamma $ is independent of $B$ (see, e.g., Eq. (2.9) of Ref.
\cite{KukushikUFN1988}). Then even the magnetic-field dependence of the
Dingle factor is different from the 3D case:
\begin{equation}
R_{\mathrm{DG}}\left( k\right) =\sqrt{\pi /2}\exp \left[ -\mathrm{const}%
\cdot k^{2}/B_{z}^{2}\right] .  \label{RDG1}
\end{equation}%
For a short-range impurity potential $d\ll l_{B}$, one obtains the
white-noise correlator $Q\left( \mathbf{r}\right) \approx \mathrm{const}%
\cdot \delta \left( \mathbf{r}\right) $. 
Then the dependence of the level width on magnetic field, in a strong field,
at $N_{LL}\gg 1$, $\Gamma \propto \sqrt{B_{z}}$ is in agreement with Refs.
\cite{Ando,Brezin}. The Dingle factor (\ref{RDG}) in this case has a similar
to the 3D case magnetic-field dependence, but a stronger damping of higher
harmonics:
\begin{equation}
R_{\mathrm{DG}}^{\ast }\left( k\right) =\sqrt{\pi /2}\exp \left[ -\mathrm{%
const}\cdot k^{2}/B_{z}\right] .  \label{RDG2}
\end{equation}

Eqs. (\ref{RDL1}),(\ref{RDL2}),(\ref{RDG1}) and (\ref{RDG2}) suggest that
it is possible not only to distinguish experimentally between the Lorentzian
and Gaussian shapes of LLs but also to obtain information about the range of
scatterers and the physical origin of the LL broadening. For the Gaussian
shape of LLs the higher harmonics of MQO are much stronger damped than for
Lorentzian LL shape because the exponent contains $k^{2}$ instead of $k$.
The Dingle plot, i.e. the plot of the logarithm of the MQO amplitudes as a
function of inverse magnetic field, gives additional information about the
origin of the LL broadening.

\section{Experimental results and discussion}

\subsection{Nonoscillating magnetoresistance}

\begin{figure}[tb]
\includegraphics[width=0.49\textwidth]{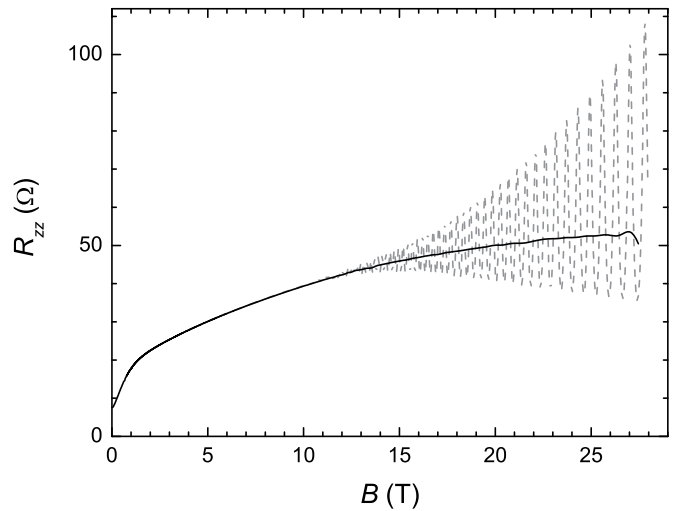}
\caption{Interlayer magnetoresistance of
$\protect\alpha $-(BEDT-TTF)$_{2}$KHg(SCN)$_{4}$
measured as a function of magnetic field perpendicular to the layers,
at $T=1.6$\;K (dashed grey line) and its monotonic component
$R_{zz}^{\mathrm{B}}(B)$ (solid black line) obtained by
filtering out the MQO, see text.}
\label{FigRzz1}
\end{figure}

Plotted in Fig. \ref{FigRzz1} are the raw data on the field dependence of
interlayer magnetoresistance $R_{zz}\left( B\right) $ of $\alpha $-(BEDT-TTF)%
$_{2}$KHg(SCN)$_{4}$, (dashed grey curve) in a field perpendicular to
layers, $\mathbf{B}\Vert z$, along with its monotonic background part $%
R_{zz}^{\mathrm{B}}$ (solid black curve), obtained by filtering out the MQO
component. Note that due to the very high amplitude of the oscillations
comparable to the monotonic background, the data should be treated in terms
of conductivity $\sigma_{zz}(B)\propto 1/R_{zz}(B)$ rather than resistivity.
Hence, for extracting the background, the as-measured resistance was first
inverted, then the oscillations were subtracted using a Fourier filter and
the result was again inverted to obtain $R_{zz}^{\mathrm{B}}(B)$ shown in
Fig. \ref{FigRzz1}.

The theory\cite{WIPRB2011,WIJETP2011,WIFNT2011} predicts that the background
magnetoresistance changes proportional to $\sqrt{B}$ in the weakly coherent
regime when $\omega _{c}\tau \gg 1$, see Eq. (\ref{sSS}) above.
\begin{figure}[tb]
\includegraphics[width=0.49\textwidth]{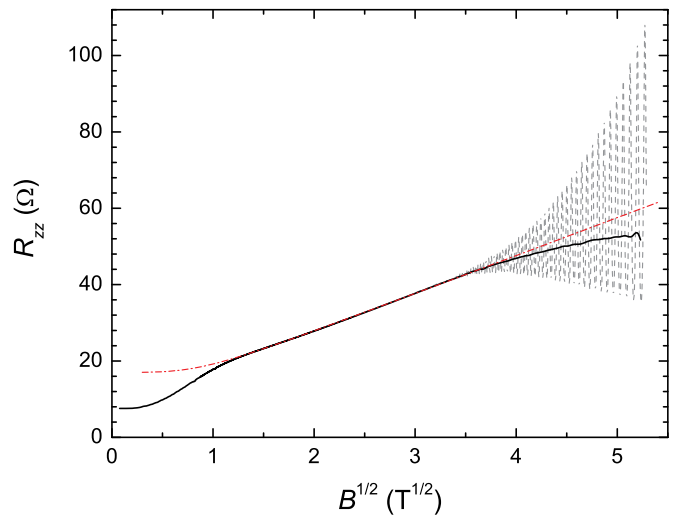}
\caption{(Color online) The same data as in Fig. \protect\ref{FigRzz1}
plotted versus $\protect\sqrt{B}$ (dashed grey and solid black lines). At
fields between 1.5 and 16\;T the magnetoresistance is linear in this scale.
The fit to Eq. (\protect\ref{sSS}) in this field range (dashed-dotted red
line) yields the transport scattering time $\protect\tau =4.3$\;ps.}
\label{FigRzz2}
\end{figure}
To compare this prediction with the observed dependence $R_{zz}\left(
B\right) $, in Fig. \ref{FigRzz2} we plot the data on $R_{zz}$ as function
of $\sqrt{B}$. From this plot one can see that background magnetoresistance
is indeed perfectly linear in this scale in the range
$1.5\mathrm{\;T}<B<16$\;T. One can fit the data in this range by
modelling the resistance as a sum of a term
$\bar{R}_{zz}(B) \propto 1/\bar{\sigma}_{zz}(B)$, determined
by Eq. (\ref{sSS}), and a field-independent term comparable to
$\bar{R}_{zz}(B=0)$. The fit shown as a dashed-dotted line in
Fig. \ref{FigRzz2} yields an estimation for the zero-field
scattering time $\tau =4.3$\;ps (using $\alpha =2$). The $B$-independent
term included in the fit appears due to
scattering on dislocations and/or phonons, which does not depend on
magnetic field and contributes to the total scattering rate $1/\tau$.

At fields below 1.5\;T
the strong-field criterion is not fulfilled for this crystal, which leads to
a deviation from the linear $\sqrt{B}$-dependence. Additionally, one has to
take into account the influence of carriers on the quasi-1D part of the
Fermi surface contributing about the same density of states as the quasi-2D
carriers considered here. The part of $\sigma _{zz}$ originating from the
quasi-1D Fermi surface rapidly (approximately quadratically) decreases with
increasing field at all field orientations except the vicinity of
commensurate directions, when the field is aligned along one of the crystal
lattice translation vectors (so-called Lebed magic angles)
\cite{mk04,osad92,lebe03}. Due to the low crystal symmetry of
$\alpha $-(BEDT-TTF)$_{2}$KHg(SCN)$_{4}$, the nearest commensurate direction
is considerably, by
$\approx 13^{\circ }$, tilted away from the $z$-axis.\cite{rous96}
Therefore, the contribution of quasi-1D carriers to $\sigma _{zz}$ is
strongly suppressed under a magnetic field $B\gtrsim B_{0}=\hbar /(e\tau
v_{F}a_{z}\tan 13^{\circ })$ applied perpendicular to layers. Substituting
the scattering time $\tau =4.3$\;ps , Fermi velocity on the quasi-1D Fermi
sheets\cite{mk09} $v_{F}=1.2\times 10^{5}$\;m/s, and the interlayer lattice
parameter\cite{rous96} $a_{z}=2.0$\;nm, we estimate $B_{0}\approx
2.7$\;T. Therefore, we attribute the steeper slope of the magnetoresistance
observed at low fields with the "freezing-out" of quasi-1D carriers. At
fields above $\sim 2$\;T the conductivity is believed to be dominated by the
carriers on the quasi-2D Fermi surface.

At fields $B>16$\;T, when the amplitude
of the oscillations becomes of the same order as the background
component $R_{zz}^{\mathrm{B}}(B)$,
the terms quadratic in the amplitude of MQO give
an additional contribution to the monotonic part of conductivity in
a way similar to that described by Eqs. (19),(21) of Ref. \cite{Shub}.
This additional contribution can be estimated as
$\Delta \sigma _{zz}\propto R_{D\ast }^{2}\approx
\exp \left( -2\pi /\omega _{c}\tau \right) $,
where the Dingle factor $R_{D\ast }$ is determined by only
short-range scattering. Substituting $\tau =4.3$\;ps and
the effective electron mass $m^{\ast }=1.3$ we obtain
$\Delta \sigma _{zz}\propto \exp \left( -B_{\ast }/B\right) $, where
$B_{\ast }\approx 11$\;T.
This explains the small deviation of the background resistivity from the
linear dependence at $B>16$\;T in Fig. \ref{FigRzz2}. Thus, in the whole,
the data in
Fig. \ref{FigRzz2} are considered as a firm evidence of the weakly-coherent
interlayer transport regime in this compound.

\subsection{Field-induced crossover in magnetic quantum oscillations}

\begin{figure}[tb]
\includegraphics[width=0.49\textwidth]{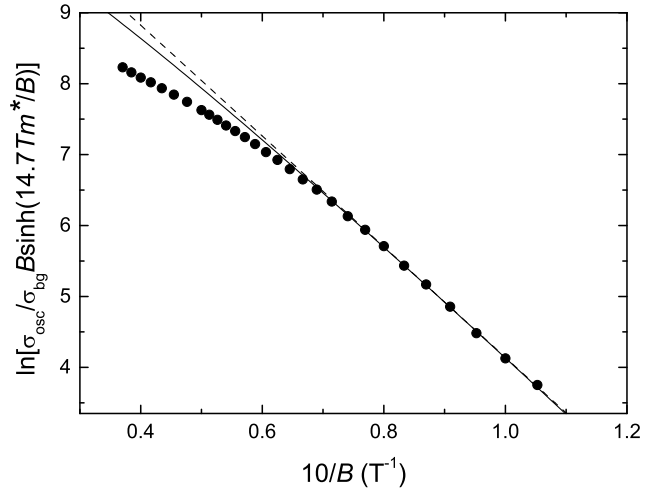}
\caption{Dingle plot, i.e. the logarithm of the MQO amplitude, divided by
the temperature damping factor $R_{T}$, as a function of inverse magnetic
field $1/B$. The dashed line is a fit of the data at fields below 12\ T to
Eq. (\protect\ref{RDL}) with field-independent $\Gamma /k_{B}=12.9$\ K, and
the solid line is the fit to Eq. (\protect\ref{RDLM}).}
\label{FigDingle1}
\end{figure}

\begin{figure}[tb]
\includegraphics[width=0.49\textwidth]{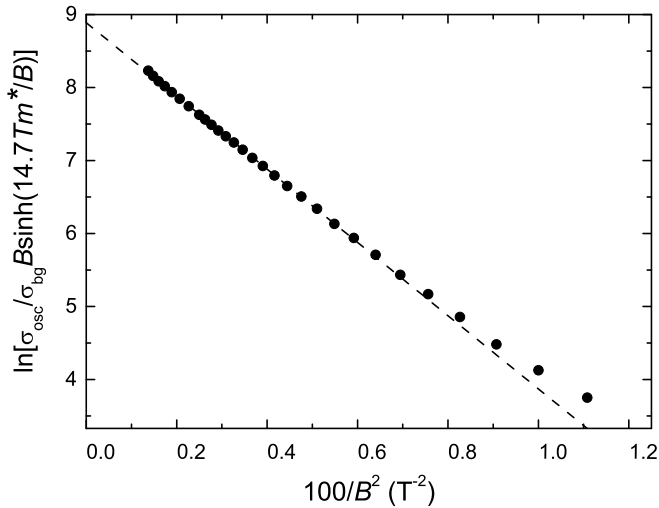}
\caption{The same data as in Fig. \protect\ref{FigDingle1} plotted against
$1/B^2$. For fields $B>12$\;T the data is nicely fitted by a straight
(dashed) line in agreement with Eq. (\protect\ref{RDG1}), implying a
Gaussian LL broadening with the halfwidth $\Gamma/k_B = 10.5$\;K determined
by a long-range scattering potential.}
\label{FigDingle2}
\end{figure}

Fig. \ref{FigDingle1} shows the Dingle plot for the first harmonic of MQO.
One can see that, contrary to the predictions of the 3D theory of
MQO, this plot is not linear in high magnetic field. This excludes the
theoretical possibilities, leading to the Dingle factors given by Eqs.
(\ref{RDL1}),(\ref{RDL2}) and (\ref{RDG2}). As was argued in Section II,
and follows from the comparison between the dashed and
solid lines in Fig. \ref{FigDingle1}, the difference between Eqs. (\ref{RDL})
and (\ref{RDLM}) on the Dingle plot is negligible and cannot explain the
observed deviation from the linear behavior. On the other
hand, the same logarithm of the MQO Dingle factor plotted as function of $%
1/B^{2}$ gives a very nice linear dependence (see Fig. \ref{FigDingle2}) at
field $B>12$\ T, which supports the scenario represented by Eq. (\ref{RDG1}%
). The LL width $\Gamma $ for $B>12$\ T is field-independent, suggesting
that the main contribution to the LL broadening comes from the long-range
random potential, which changes on the length $d\gg l_{B}$ and gives local
variations of the Fermi energy. This long-range potential only damps the MQO
but it does not affect the background (averaged over MQO) conductivity
because it does not produce a significant electron scattering and relaxation
of electron momentum. This situation is similar to that observed in Ref.
\cite{SO}, where the long-range disorder potential only damped the fast MQO
but did not damp the slow oscillations of magnetoresistance. Hence, Eq. (\ref%
{sSS}) remains valid, because $\Gamma =\hbar /2\tau $ is determined by
short-range impurities and increases in strong magnetic field $\propto \sqrt{%
B}$. The fact that LL broadening is Gaussian is also very important: it
means that electron dynamics in $\alpha $-(BEDT-TTF)$_{2}$KHg(SCN)$_{4}$
under a strong field is indeed substantially different from that in 3D
metals where the impurity scattering leads to a finite electron lifetime and
produces the Lorentzian level broadening.

At fields $B<12$\;T the dependence in Fig. \ref{FigDingle2} deviates from
linear, suggesting a crossover from the high-field Gaussian LL shape to
another shape at lower field, probably, to the Lorentzian shape with a
field-independent width $\Gamma $. The linear fit of the Dingle plot in
Fig. \ref{FigDingle1} at $9.5\mathrm{\;T}<B<12$\;T gives the LL width
$\Gamma/k_B =\pi T_{D}\approx 12.9$\;K, which is $6$ times greater than
one would naively expect from the transport relaxation time
$\tau \approx 4.3$\;ps determined by short-range scattering. This means
that the LL broadening is determined by the long-range disorder potential,
which does not produce electron scattering. Fitting of the high-field Dingle
factor in Fig. \ref{FigDingle2} by Eq. (\ref{RDG}) gives a comparable LL
width $\Gamma/k_B \approx 10.5$\;K.

Now we use the obtained values of $\Gamma $ to analyze the damping of MQO
harmonics and to compare the theoretical predictions for the harmonic
amplitudes for both LL shapes with the experimental data. We remind that,
taking into account the large amplitude of the oscillations, the analysis is
performed for inverse resistance $1/R_{zz}(B)\propto \sigma _{zz}(B)$.
\begin{figure}[t]
\includegraphics[width=0.49\textwidth]{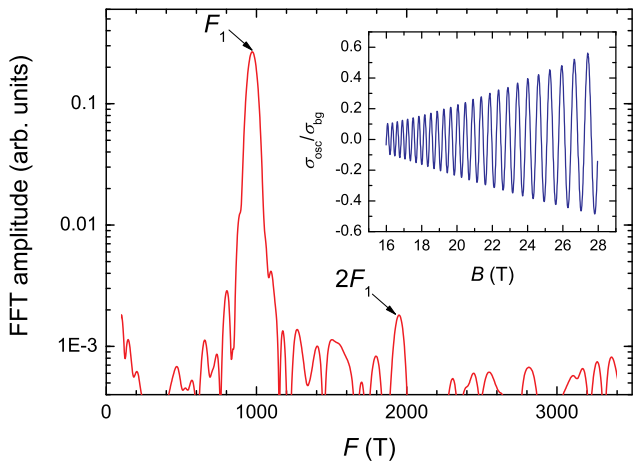}
\caption{(Color online) FFT spectrum of the oscillations in the interlayer
conductivity in the field range $16\mathrm{\;T}<B<28$\ T, as shown in the
inset.}
\label{FigOsc}
\end{figure}
Fig. \ref{FigOsc} shows the fast Fourier transform (FFT) of the oscillatory
component of inverse resistance normalized to the field-dependent
nonoscillating background. The data is taken in the field window
$16\mathrm{\;T}<B<28$\ T, as shown in the inset in Fig. \ref{FigOsc}. One
can see that
the Fourier spectrum is almost completely dominated by one fundamental
harmonic. The amplitude of the second harmonic only slightly exceeds the
noise, while the third harmonic is not resolved within the present accuracy.
The ratio of the FFT amplitudes of the first and second harmonics averaged
over the given field window is $A_{2}/A_{1}\approx 150$, while the first
harmonic amplitude normalized to the nonoscillating background increases
from $A_{1}\approx 0.1$ at $B=16$\ T to $A_{1}\approx 0.5$ at $B=28$\ T (see
the inset in Fig. \ref{FigOsc}). For the analysis we take the average (in the
$1/B$-scale) value $B_{a}=20.4$\ T, where the experimentally obtained
normalized amplitudes are $A_{\mathrm{1,exp}}\approx 0.25$ and
$A_{\mathrm{2,exp}}\approx 1.7\times 10^{-3}$. The temperature in the experiment
is $T\approx 1.6$\ K, and the electron effective mass at pressure $6$\ kbar is
$m^{\ast }\approx 1.3$. This gives $X\equiv 2\pi ^{2}k_{B}T/\hbar \omega
_{c}\approx 1.50$ at $B_{a}=20.4$\ T, and the temperature damping factors of
the first and second harmonics are $R_{T}\left( 1\right) =0.70$ and
$R_{T}\left( 2\right) =0.30$. The experimental error bar in determination of
the electron effective mass gives the possible errors in the temperature
damping factor $\sim 5\%$ and $\sim 15\%$ for the first and second harmonic,
respectively. The spin reduction factor $R_{\mathrm{S}}$ can be evaluated
from spin-zeros experiments. So far, such experiments have been done for
$\alpha $-(BEDT-TTF)$_{2}$KHg(SCN)$_{4}$ only at ambient pressure, (at high
magnetic fields, where the CDW gap is strongly suppressed),
yielding\cite{gfactorSasaki,gfactorHarrison} $gm^{\ast }=3.65\pm 0.02$. Making a
correction to the pressure-dependent effective mass, which changes from
$m^{\ast }\approx 2.0$ at ambient pressure to 1.3 at 6\ kbar, and assuming a
pressure-independent $g$-factor we substitute $gm^{\ast }=2.37$ in the spin
reduction factor to obtain a rough estimate $R_{\mathrm{S}}(1)\simeq 0.8$ and
$R_{\mathrm{S}}(2)\simeq 0.4$.

For the Lorentzian LL shape with field-independent $\Gamma/k_B = 12.9$\;K
one obtains from Eq. (\ref{RDL}) at $B_{a}=20.4$\;T the Dingle factors
$R_{\mathrm{DL}}\left( 1 \right) \approx 0.022$ and $R_{\mathrm{DL}}\left(
2\right) \approx 0.00047$. The predicted harmonic amplitudes for the
Lorentzian LL shape are $A_{\mathrm{1,th}}=R_{\mathrm{DL}}\left( 1\right)
R_{T}\left( 1\right) R_{\mathrm{S}}(1)\approx 0.022\times 0.70\times
0.8=0.012 $ and $A_{\mathrm{2,th}}=R_{\mathrm{DL}}\left( 2\right)
R_{T}\left( 2\right) R_{\mathrm{S}}(2)\approx 0.00047\times 0.30\times 0.4=
5.6\times 10^{-5}$, which is much smaller than the experimental values.
The smaller value $\Gamma/k_B = 10.5$\;K obtained for Gaussian LL shape
gives the Dingles factors $R_{\mathrm{DL}}\left( 1\right) \approx 0.044$ and
$R_{\mathrm{DL}}\left( 2\right) \approx 0.0020$, and the harmonic amplitude
$A_{\mathrm{1,th}}= 0.025$ and $A_{\mathrm{2,th}}=0.00024$, which still by an
order of magnitude differs from the experimental data. Thus, the observed
harmonic amplitudes are inconsistent with the traditional 3D Dingle factor,
corresponding to the Lorentzian LL shape.

For the Gaussian LL shape with field-independent $\Gamma /k_{B}=10.5$\ K one
obtains from Eq. (\ref{RDG}) at $B_{a}=20.4$\ T the Dingle factors $R_{
\mathrm{DG}}\left( 1\right) \approx 0.37$ and $R_{\mathrm{DG}}\left(
2\right) \approx 0.0099$. Then the calculated harmonic amplitudes for the
Gaussian LL shape are $A_{\mathrm{1,th}}=R_{\mathrm{DG}}\left( 1\right)
R_{T}\left( 1\right) R_{\mathrm{S}}(1)\approx 0.37\times 0.70\times 0.8=0.22$
and $A_{\mathrm{2,th}}=R_{\mathrm{DG}}\left( 2\right) R_{T}\left( 2\right)
R_{\mathrm{S}}(2)\approx 0.0099\times 0.30\times 0.4=0.0012$, which nicely
agrees the experimental values $A_{\mathrm{1,exp}}\approx 0.25$ and $A_{
\mathrm{2,exp}}\approx 0.0017$. This analysis gives an additional
substantiation that the standard 3D formulas for electron scattering are not
applicable at high magnetic fields. The observed electron interaction with a
disorder potential corresponds to the 2D theoretical models rather than to
the 3D electron dynamics. This is the origin of the new features of
magnetoresistance in the weakly coherent limit.

Note that the crossover in the $B$-dependent MQO amplitude occurs at
considerably higher fields than the crossover to the $\sqrt{B}$-behavior
of the background resistance $R_{zz}^{\mathrm{B}}(B)$. This is because
the strong-field criteria, which are formally similar for both
crossovers, $\hbar\omega_c/\Gamma = 2\omega_c\tau \gg 1$, in fact, involve
different scattering parameters $\Gamma$ (or, equivalently, $1/\tau$):
as shown above, the short-range impurity scattering rate, which determines
$R_{zz}^{\mathrm{B}}(B)$, is considerably lower than that coming from the
long-range disorder and dominating in the LL broadening.

\section{Concluding remarks}

In conclusion, we note that the above dimensional
coherent -- weakly-coherent
crossover is, strictly speaking, not a complete 3D$\rightarrow $2D
transition, because the interlayer hopping time $\tau _{h}$ is still much
shorter than the electron phase decoherence time $\tau _{\phi }$. This fact
is crucial for the Anderson localization and, hence, for the temperature
dependence of conductivity $\sigma \left( T\right) $. If electrons are
localized, the resistivity increases with decreasing $T$. Most of organic
metals, including the present compound, have metallic $\sigma \left(
T\right) $ at low temperatures, which excludes electron localization, at
least on the length of the sample size.\cite{Comment1} Therefore, the
quantum Hall effect (QHE), which requires electron localization, is not
observable in bulk layered metals.\cite{CommentFisdw}

Indeed, the 2D electron localization length in a strong magnetic
field,\cite{Fogler1997} $\xi \sim R_{c}\exp (\pi ^{2}g_{0}^{2})$, where the
dimensionless conductivity $g_{0}=\left( h/e^{2}\right) \sigma _{xx}\approx
\left( 2N_{LL}+1\right) /\pi $ in the MQO maxima,\cite{Ando01} $N_{LL}$ is
the number of occupied Landau levels, and the cyclotron (Larmor) radius
\begin{equation}
R_{c}=\hbar k_{F}/eB=k_{F}l_{B}^{2}=\left( 2N_{LL}+1\right) /k_{F},
\label{Rc}
\end{equation}
and $k_{F}$ is the inplane Fermi momentum. For the electron localization
and the QHE to take place, the electrons must travel (diffusively) on the
distance $\xi $ before loosing the phase or jumping to the next layer even
at the DoS maxima. This gives the condition $\xi ^{2}/D<\min \left\{ \tau
_{\phi },\tau _{h}\right\} $, with $D\approx R_{c}^{2}/2\tau $ being the 2D
diffusion coefficient. This condition, equivalent to $\min \left\{ \tau
_{\phi },\tau _{h}\right\} /\tau \gtrsim \exp \left( 4N_{LL}^{2}\right) $,
is too strict to be fulfilled in bulk layered metals, where $N_{LL}\gg 1$.%
\cite{Comment2} Hence, the organic metals as well as all the other known
bulk conductors with a quasi-2D electronic structure can only have an
\textit{incomplete} 3D$\rightarrow $2D crossover.

At increasing temperature, the conductivity due to direct tunneling
decreases and other conduction mechanisms associated, e.g., with small
polarons \cite{Lundin2003,Ho} or resonant impurity tunneling
\cite{Abrikosov1999,Maslov,Incoh2009} may come into play. This may lead to
a crossover from a low-temperature metallic to a high-temperature, apparently,
nonmetallic $T$-dependence of $\rho _{zz}$ which was reported for
various layered materials.

To summarize, we have proposed and substantiated the field-induced
dimensional crossover in strongly anisotropic quasi-2D layered compounds. In
high magnetic field, when $\omega _{c}>t_{z},1/\tau $, a qualitatively new
\textit{weakly coherent} regime of interlayer magnetotransport emerge. In
this regime the monotonic parts of interlayer magnetoresistance $R_{zz}(B)$
and the harmonic damping of MQO show the behavior, completely different from
that predicted by the traditional 3D theory generalized to quasi-2D
case.\cite{Shub,ChampelMineev,MosesMcKenzie1999} The experimental results on
$R_{zz}(B)$ agree very well with the new theoretical predictions and provide
valuable information about scattering processes in the crystal.

\section{Acknowledgments}

We are grateful to H. M\"uller and N. D. Kushch for providing
high-quality crystals of $\alpha$-(BEDT-TTF)$_{2}$KHg(SCN)$_{4}$.
The measurements in magnetic fields up to 28\;T were carried out
at the Laboratoire National des Champs Magn\'{e}tiques Intenses,
Grenoble, France. The work was supported by EuroMagNET II under
the EU contract 228043, by DFG grant No. Bi 340/3\_1 and by the
Federal Agency of Science and Innovations of Russian Federation
under Contract No. 14.740.11.0911.


\begin{thebibliography}{99}
\bibitem{commentFISDW} Here we don't consider magnetic-field-induced density
wave transitions observed in quasi-one-dimensional organic metals.\cite%
{lebed84,B1}
\bibitem{Comment1} Note that the nonmetallic $\sigma \left( T\right) $ does
not necessarily mean the presence of electron localization.
\end{thebibliography}
\end{document}